\documentclass[twocolumn]{aastex61}

\usepackage{amsmath,amstext}
\usepackage[T1]{fontenc}
\usepackage{apjfonts} 
\usepackage{hyperref}
\usepackage[figure,figure*]{hypcap}
\usepackage{hyperref}
\hypersetup{linkcolor=red,citecolor=green,filecolor=cyan,urlcolor=magenta}
\usepackage{epsfig}
\usepackage{amssymb}
\usepackage{graphicx}
\usepackage{bm}
\usepackage{verbatim}
\usepackage{cancel}
\usepackage{mathtools}
\usepackage{color}
\usepackage{accents}
\usepackage{gensymb}

\newcommand{\change}[1]{\textcolor{black}{#1}}

\newcommand{\dd}[2]{\frac{d #1}{d #2}}

\newcommand{\pd}[2]{\frac{\partial#1}{\partial#2}}

\newcommand{\grad}[0]{\vect{\nabla}}
\newcommand{\vect}[1]{\ensuremath{\underaccent{\bar}{\bf{#1}}}}

\newcommand{\slashsigma}[0]{{\sigma_1} }

\newcommand{\deltastar}[0]{\Delta^{\!\star}}

\newcommand{\smallparenth}[1]{\left(\kern-0.5pt #1\kern-0.5pt \right)}
\newcommand{\smallabs}[1]{\left|\kern-1pt #1\kern-1pt \right|}
\newcommand{\functionof}[2]{#1\kern-1pt\smallparenth{#2}}
\newcommand{\dotprod}[2]{#1\!\cdot\!#2}

\newcommand{\leedot}[0]{\!\cdot\!}

\DeclareRobustCommand{\rchi}{{\mathpalette\irchi\relax}}
\newcommand{\irchi}[2]{\raisebox{0.7\depth}{$#1\chi$}} % inner command, used by \rchi

\published{2019 January 4}
\submitjournal{ApJ}

\shorttitle{A Model of Solar Equilibrium}
\shortauthors{Gunderson et al.}

\begin{document}

\title{A Model of Solar Equilibrium:\\The Hydrodynamic Limit}

\author{L. M. Gunderson}
\affiliation{Center for Heliophysics, Department of Astrophysical Sciences and Princeton Plasma Physics Laboratory, Princeton University, Princeton, NJ 08540, USA}
\author{A. Bhattacharjee}
\affiliation{Center for Heliophysics, Department of Astrophysical Sciences and Princeton Plasma Physics Laboratory, Princeton University, Princeton, NJ 08540, USA}

\begin{abstract}
\noindent
Helioseismology has revealed the internal density and rotation profiles of the Sun.  Yet, knowledge of its magnetic fields and meridional circulation is confined much closer to the surface, and latitudinal entropy gradients are below detectable limits.  While numerical simulations can offer insight into the interior dynamics and help identify which ingredients are necessary to reproduce particular observations, some features of the Sun can be understood analytically from an equilibrium perspective.  Examples of such features include:  the 1D density profile arising from steady-state energy transport from the core to the surface, and the tilting of isorotation contours in the convection zone (CZ) due to baroclinic forcing.  To help identify which features can be explained by equilibrium, we propose analyzing stationary axisymmetric ideal magnetohydrodynamic flows in the solar regime.  By prescribing an appropriate entropy profile at the surface, we recover a rotation profile that reasonably matches observations in the bulk of the CZ.  Additionally, by including the effects of poloidal flow, we reproduce a feature that is reminiscent of the near surface shear layer.  However, no tachocline-like feature is seen in hydrodynamic equilibrium, suggesting the importance of either dynamics or magnetic fields in its description.
\end{abstract}


\section{Introduction}
Helioseismology has revealed the rich structure of the internal rotation of the Sun \citep{christensen2002helioseismology}.  Notably, while the radiative interior rotates uniformly, the transition to differential rotation in the convection zone (CZ) occurs in a narrow shear layer known as the tachocline.  In addition, rotation decreases in the outer ${\rm\sim~\!\!5\%}$ of the Sun's radius, forming another region known as the near surface shear layer (NSSL).  A fundamental goal of solar physics is to understand the reasons for this rotation profile and its implications for stellar models.

In the bulk of the CZ (i.e.,~away from its boundaries), the isorotation contours are tilted radially.  This deviation from the cylindrical rotation of Taylor--Proudman balance requires a source of azimuthal vorticity, and the commonly posited source is baroclinic forcing, as it requires the poles to be only ${\rm\sim~\!\!10~K}$ warmer \change{than the equator} to match observations.  Although this gradient cannot be directly measured, some simulations have managed to obtain solar-like tilted contours through this effect, for example, the 2D mean-field hydrodynamic (HD) models of \citet{kitchatinov1995differential} and \citet{rempel2005solar}.  The 3D simulations of \citet{miesch2006solar} also found a similar tilting of contours, albeit with an imposed entropy gradient at the base of the CZ.  However, most 3D simulations have found rotation profiles closer to the Taylor--Proudman state than observations \citep{featherstone2015meridional}.

The first evidence for the NSSL was the observation that emerging active regions rotate faster than the surrounding photosphere \citep{foukal1972magnetic}.   The usual explanation for its existence is the radial mixing of angular momentum due to strong convection in the outer layers \citep{choudhuri1995solar,miesch2011gyroscopic}.  High-resolution simulations of convection have recovered similar profiles \citep{hotta2014high}.  

The meridional (poloidal) flow is relatively less constrained by observations; the only conclusive observation seems to be a ${\rm\sim~\!\!20~m~s^{- 1}}$ poleward surface flow at \mbox{mid-latitudes} \citep{rajaguru2015meridional}.  As the meridional flow is not thought to penetrate the radiative interior \citep{gilman2004limits}, this observation is often extrapolated to suggest a single recirculating cell in each of the north and south hemispheres, with an equatorward return flow near the base of the CZ.  Mean-field models of the solar cycle (e.g.,~\mbox{Babcock-Leighton)} frequently use such a profile \citep{dikpati1999babcock}.  However, in \citet{zhao2013detection}, local helioseismological measurements were found to implicate an equatorward return flow closer to the surface, indicating a more complex topology than previously assumed.  The \mbox{large-scale} magnetohydrodynamic (MHD) simulations in \citet{passos2015meridional,passos2016new} corroborated this picture, suggesting a \mbox{multicelled} pattern with strong axial alignment near the equator.

In an attempt to find simple closed form solutions for the CZ rotation, \citet{balbus2009simple} found strikingly good agreement with observations using a very simple model.  Its two ingredients are: differential rotation in thermal wind balance (TWB; i.e.,~baroclinic forcing in the hydrostatic limit), and the assumption that the aspherical part of the entropy is constant along isorotation contours.  Indeed, when differential rotation is observed in solar simulations, it appears to be in dominant balance with a latitudinal entropy gradient throughout the bulk of the CZ \citep{passos2017meridional}.  The second assumption, however, is more questionable, although possible mechanisms for this alignment were posited in \citet{balbus2009differential,balbus2012global}, \citet{balbus2010tachocline}, and \citet{balbus2012stability}, including a marginally stable magnetobaroclinic mode, and a dynamic alignment of convective structures due to rotational shear.

Our work is motivated by the observation that a similar alignment of contours naturally appears in the solution to axisymmetric ideal MHD equilibrium with flows. The equation governing this system, generalizing the standard Grad--Shafranov (G--S) approach for static plasmas, was derived by several authors (e.g.,~\citet{hameiri1983equilibrium,lovelace1986theory,goedbloed1997stationary}). In this framework, certain quantities are \mbox{flux/stream} functions, i.e.,~they are constant along the surfaces traced out by the poloidal field and flow.  For the assumed closure of adiabatic flows, entropy is one such stream function.  In the limit where the poloidal field is strong compared to poloidal flow, toroidal rotation is also a stream function.  This would \textit{seem} to imply Balbus's assumption of aligned entropy and isorotation contours.  However, observations of the mean field and flow at the solar surface suggest that the CZ is in the opposite limit, where the poloidal flow dominates over the poloidal field.  In this limit, angular momentum replaces rotation as a stream function, implying that the natural equilibrium prescription is for entropy to align instead with angular momentum.  This case was also considered in \citet{balbus2009simple}, but was discarded as initial results did not qualitatively match helioseismic observations.  However, we demonstrate that the freedom in choosing the form of the stream functions in the generalized G--S framework allows the theoretical model to match more closely with observations.

Indeed, given the variety of competing models of solar behavior (e.g.,~\mbox{steady-state/}\mbox{mean-field/}convection, \mbox{HD/MHD}, \mbox{2D/3D}, boundary conditions, etc.), it is essential to have a way to make direct comparisons and to determine which elements are required to explain different features.  For example, do the gross features of rotation and circulation in the CZ require magnetic fields \citep{passos2015meridional,passos2016new}, or does an HD explanation suffice?  How important are convective motions for the NSSL \citep{miesch2011gyroscopic,hotta2014high}?  Can the tachocline be understood from an equilibrium perspective, or must turbulence be invoked \citep{spiegel1992solar}?  Does the tachocline require magnetic fields \citep{gough1998inevitability}, and if so, can it exist without a solar cycle \citep{barnabe2017confinement}?  To what extent can torsional oscillations be understood as a series of quasi-static equilibria \citep{rempel2007origin,guerrero2016understanding}?  To obtain answers to some of these questions, which have been considered extensively in the literature, we propose a sort of \mbox{``null model,''} using the G--S framework to study solar-relevant equilibrium states.  This approach would help determine which features can be described by equilibrium, and which are necessarily dynamical in origin.  In this paper, we focus on the CZ and NSSL in the HD limit.

In Section~\ref{sec:equations}, we review the equations describing stationary axisymmetric ideal MHD flows and its HD limit.  In \mbox{Section~\ref{sec:balbus},} we show how a \mbox{Balbus-like} ansatz is naturally included in the G--S framework and compare the two models.  Effects of the poloidal flow are presented in \mbox{Section~\ref{sec:poloidaleffects}}, including a surface feature qualitatively similar to the NSSL.  In \mbox{Section~\ref{sec:conclusions}}, we discuss potential applications, limitations, and extensions of our model.

\section{Grad--Shafranov Equation with Flows}
\label{sec:equations}
Our starting point is stationary, axisymmetric, ideal MHD:
	\begin{align}
	\grad \cdot \left ( \rho \vect{v} \right ) &= 0, \quad \grad \cdot \vect{B} = 0, \label{masseq} \\
	\rho\vect{v}\cdot\grad \vect{v}+\grad p + \rho \grad G &= \left ( \grad \times \vect{B} \right ) \times \vect{B}, \label{momentumeq}\\
	\grad \times \left ( \vect{B} \times \vect{v} \right ) &= 0, \label{faraday} \\
	\vect{v}\cdot\grad \sigma &= 0, \quad \sigma \equiv \ln \left (p /\rho^{\gamma}\right), \label{entropyeq} \\
	\pd{}{\phi} &= 0,
	\end{align}
where $\rho$, $\vect{v}$, $\vect{B}$, $p$, $\sigma$, and $G$ are the density, fluid velocity, magnetic field, pressure, entropy, and external potential, respectively, and we have absorbed the permeability into the units of the magnetic field.  We use $\left ( r, \theta, \phi \right )$ for spherical and $\left ( \lambda, z, \phi \right )$ for cylindrical coordinates, with $\lambda = r\sin \theta $, \mbox{$z = r\cos \theta$,} and $\phi$ as the longitudinal coordinate.

The solution to these equations is known as the generalized G--S equation and has found a variety of applications.  The magnetostatic version was originally derived to model thermonuclear confinement devices \citep{grad1958hydromagnetic}, and is still often used for tokamak equilibrium reconstruction.  The observation of a toroidal rotation in these experiments motivated the inclusion of flows, and the numerical solutions of \citet{guazzotto2004numerical} suggested the possible importance of poloidal flows for internal transport barriers.  More recently, it has seen a variety of astrophysical applications.  When applied to stellar winds, it can be seen as a 2D magnetic generalization of the Parker wind model \citep{okamoto1975magnetic,heinemann1978axisymmetric,sakurai1985magnetic}.  It can also be used to reconstruct 2D coherent structures within the solar wind and magnetosphere using only measurements from a single spacecraft \citep{sonnerup2006grad}.  Various generalizations have further extended its applications: relativity has been included for modeling accretion disks \citep{lovelace1986theory,goedbloed2004unstable,beskin2004grad}, and superconductivity for neutron stars \citep{lander2009magnetic,lander2013magnetic}.  Given its success in providing plausible equilibrium models from incomplete observational data, we propose an application to another compact astrophysical object, the Sun.  

This paper is concerned with the HD limit, thus, we outline the solution in the limit of zero magnetic field.  As $\rho \vect{v}$ is divergenceless and axisymmetric, we can express the velocity as
\begin{equation}
\vect{v} = \frac{1}{\rho} \grad \phi \times \grad \rchi + v_\phi \lambda \grad \phi,
\end{equation}
hence $\rchi$ is constant in the direction of the the poloidal flow, i.e.,~it is a stream function.  

In this HD limit, the toroidal component of Equation~\eqref{momentumeq} yields
\begin{equation}
\functionof{\grad}{\lambda v_\phi} \times \grad \rchi = 0,
\end{equation}
implying that angular momentum is also a stream function, i.e.,~$L \equiv \lambda v_\phi = \functionof{L}{\rchi}$.  Likewise, Equation~\eqref{entropyeq} implies that \mbox{$\sigma = \functionof{\sigma}{\rchi}$} is a stream function.   Force balance (Equation~\eqref{momentumeq}) in the direction of the flow yields a generalized Bernoulli equation and the last stream function, $\functionof{H}{\rchi}$:
\begin{align}
\frac{1}{2} \frac{\left | \grad\rchi \right |^2}{\rho^2 \lambda^2} + \frac{1}{2} \frac{{L}^2}{\lambda^2} + \frac{\gamma\rho^{\gamma-1}}{\gamma-1} e^\sigma + G - H = 0.\label{goedbloedBernhydro} 
\end{align}
Force balance in the direction perpendicular to the poloidal flow gives the G--S PDE: %\change{FUCKING FIX THIS EQUATION IN THE PUBLISHED VERSION}
\begin{align}
&\grad \leedot \left ( \frac{\grad\rchi}{\rho \lambda^2} \right ) + \frac{\rho}{2\lambda^2} {L^2}'  + \frac{\rho^{\gamma}e^\sigma}{\gamma-1} \sigma'  - \rho H' = 0, \label{goedbloedPDEhydro}%+ \frac{\rho^2 \lambda^2 }{2} {I^2}'
\end{align}
where the primes denote partial differentiation with respect to $\rchi$.

We rewrite Equations~\eqref{goedbloedBernhydro} and \eqref{goedbloedPDEhydro} as a differential part, containing the spatial derivatives of $\rchi$, and collect the remaining terms as an algebraic part, that is:
	\begin{equation}
	\underbrace{\deltastar \!\left [ \rchi \right ]}_{\text{differential part}} \quad\!\!\!\!\!\!\! + \quad \underbrace{\functionof{F}{\rchi, \rho, \lambda, z}}_{\text{algebraic part}} \,\, = \,\, 0. \label{schematicequation}
	\end{equation}

\section{Comparison with Balbus Model}
\label{sec:balbus}

\begin{figure*}
\begin{center}
%\includegraphics*[width=6cm,height=5.1cm,angle=0]{BalbusOmegaLabelsRemoved.pdf}\includegraphics*[width=6cm,height=5.1cm,angle=0]{BalbusAngularLabelsRemovedGOOD.pdf}\\
%\hspace{-0.45cm}\includegraphics*[trim={7.8cm 4.0cm 2cm 2.4cm},clip,width=5.8cm,angle=0]{New1stPicBottomLeft.pdf}\hspace{0.2cm}\includegraphics*[trim={7.8cm 4.0cm 2cm 2.4cm},clip,width=5.8cm,angle=0]{New1stPicBottomRight.pdf}
\includegraphics*[width=12cm,angle=0]{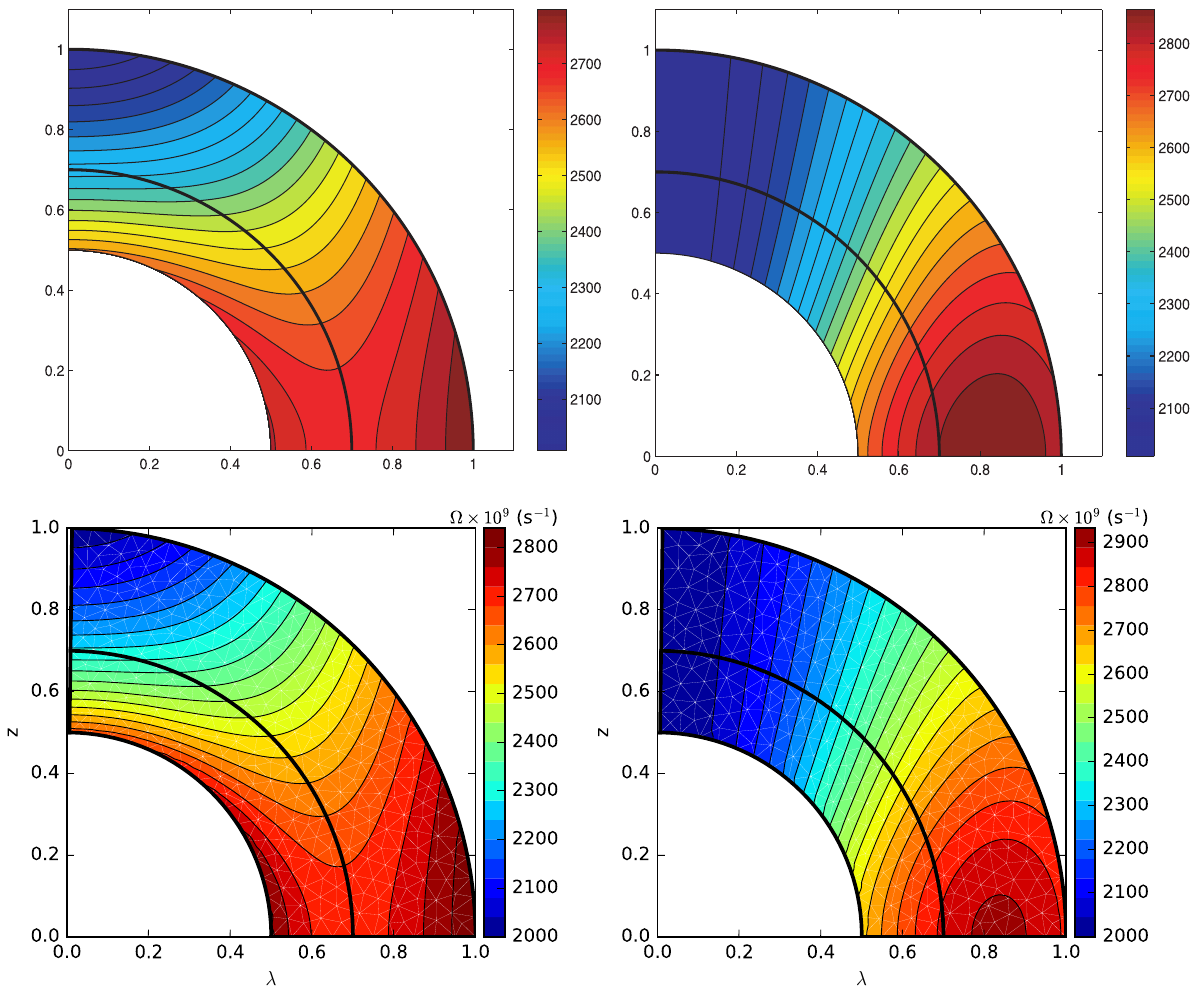}
\caption{Comparison of rotation profiles of TWB solutions in \citet{balbus2009simple} with our results from Grad--Shafranov theory.  \textbf{Top left}: TWB (with the \mbox{$\sigma$--$\Omega$} alignment) from \citet{balbus2009simple}.  \textbf{Top right}: TWB (with the \mbox{$\sigma$--$L$} alignment) from \citet{balbus2009simple}.  In both cases, rotation at the surface was chosen to match solar observations, but the functional form of the entropy is prescribed as $\sigma \propto - \Omega^2$ and $\sigma \propto - L^2 = - \Omega^2 \lambda^4$, respectively.  The gradient of the latter function occurs at a larger cylindrical radius, $\lambda$, resulting in a lack of tilting of contours at mid to high latitudes and overly tilted contours near the equator.  \textbf{Bottom}: TWB solutions from our G--S theory (i.e.,~with the \mbox{$\sigma$--$L$} alignment), \change{where rotation} and entropy at the surface are chosen to match those above.  Note that the choice of entropy alignment in the CZ matters much less than the choice of entropy profile at the surface.}
\end{center}
\label{fig:balbusrotationcomparison}
\end{figure*}

\begin{figure*}[t]
\begin{center}
\includegraphics*[trim={0cm -0.5cm 0 0},clip,width=6.2cm,angle=0]{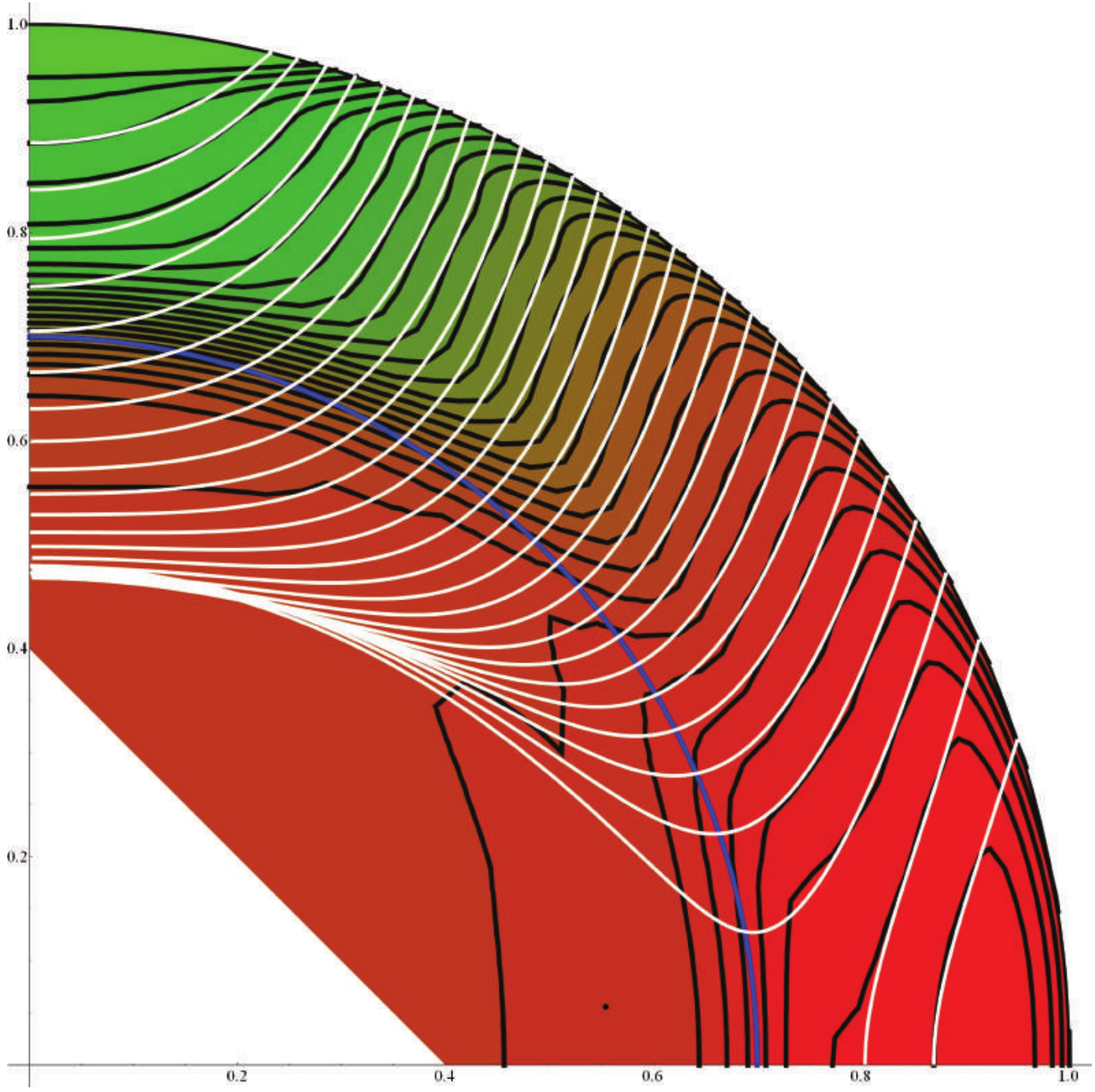}\vspace{0.1cm}
\includegraphics*[trim={4cm 0.8cm 0 0},clip,width=8.4cm,angle=0]{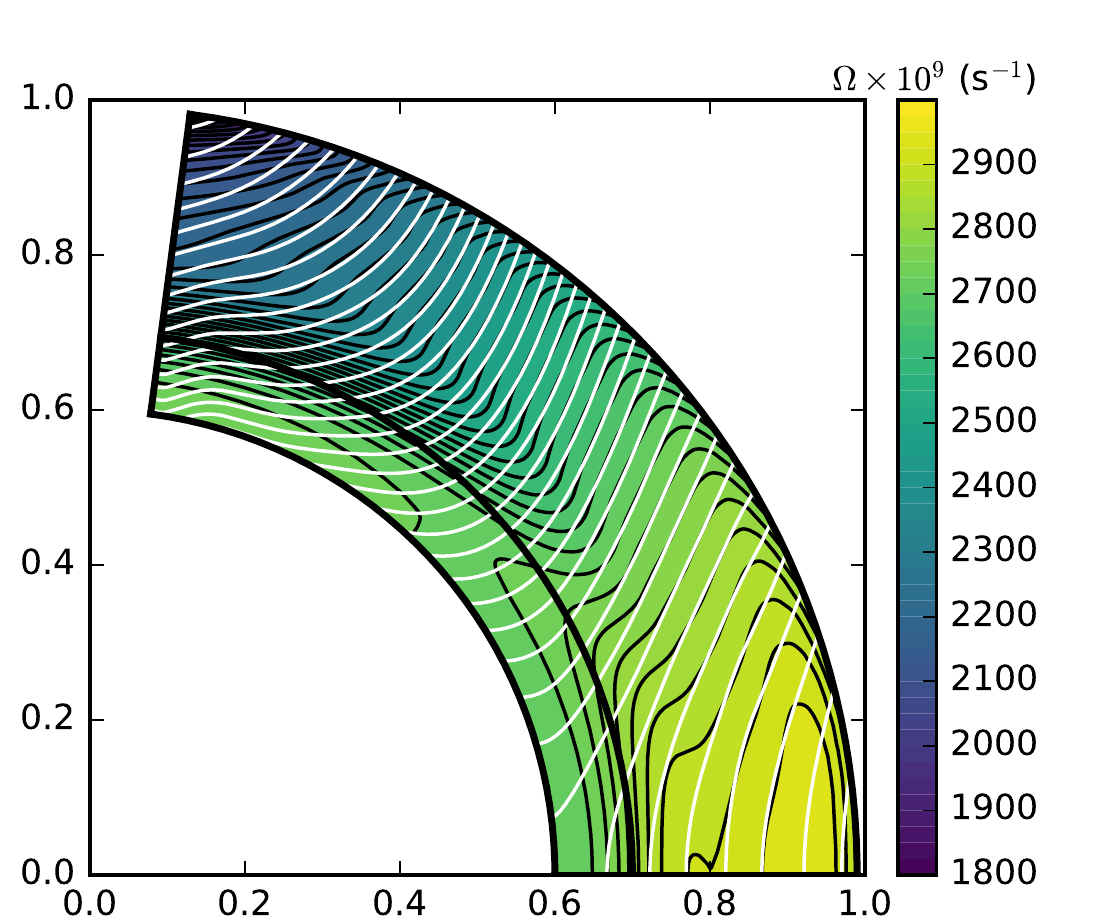}
\end{center}
\caption{Comparison of Balbus's model and our G--S model with the rotation profile deduced by helioseismology. \textbf{Left}: the best fit from \citet{balbus2009differential} in white, with helioseismic data in black.  \textbf{Right}: the best fit from our G--S model in white, with data from \citet{howe2011first} in black.  Note the similar agreement in the bulk of the CZ (away from the surface and the tachocline).}
\label{fig:helioseismologyfit}
\end{figure*}

\subsection{Thermal wind balance limit}
The first assumption in \citet{balbus2009simple} is that of TWB.  It considers a rotating fluid with a background state of gravitationally dominated hydrostatic balance.  For the case of a baroclinic fluid, the Taylor--Proudman theorem states that this rotation profile must be cylindrical, i.e.,~with isorotation contours that align with the axis.  TWB considers the case where deviations from the Taylor--Proudman profile are caused by baroclinic forcing (i.e.,~vorticity due to misaligned density and pressure gradients).

In this section, we show that we can recover the TWB equation from the HD G--S equation by neglecting the poloidal flow.  In this limit, the PDE~\eqref{goedbloedPDEhydro} and Bernoulli equation~\eqref{goedbloedBernhydro} become the following:
	\begin{align}
	&\frac{{L^2}'}{2 \lambda^2} + \frac{1}{\gamma-1} \frac{p}{\rho} \sigma'  - H' = 0, \label{reducedpde}\\
	&\frac{{L^2}}{2 \lambda^2} + \frac{\gamma }{\gamma-1} \frac{p}{\rho} + G - H = 0, \label{reducedbernoulli}
	\end{align}
where $L^2 = \Omega^2 \lambda^4$ and $\Omega \equiv \frac{v_\phi}{\lambda}$.  

We take the derivative of Equations~\eqref{reducedpde} and~\eqref{reducedbernoulli} along the stream surfaces, i.e.,~in the direction of $\vect{X} \equiv \grad\phi \times \grad\rchi$.  To simplify the expressions, note the following:
	\begin{align}
	&\dotprod{\vect{X}}{\grad \lambda^2} = 2 \pd{\rchi}{z},  \nonumber\\
	&\dotprod{\vect{X}}{\grad r} = -\frac{1}{\lambda r}\pd{\rchi}{\theta},  \nonumber\\
	&\dotprod{\vect{X}}{\grad f} = 0 \quad \forall f\in\left\{ \rchi, L, \sigma, H \right\}. \nonumber
	\end{align}
Then, we evaluate $\vect{X}\leedot\grad$ of Equation~\eqref{reducedbernoulli}:
	\begin{align}
	\frac{\gamma}{\gamma-1} \vect{X} \leedot \grad \frac{p}{\rho} &= -\vect{X} \leedot \grad \frac{{L^2}}{2 \lambda^2} - \vect{X} \leedot \grad G + \vect{X} \leedot \grad H \nonumber\\
	&= \frac{{L^2}}{\lambda^4} \pd{\rchi}{z} + \frac{g}{\lambda r} \pd{\rchi}{\theta}, \label{xDotGrad14}
	\end{align}
where $\functionof{g}{r} = \dd{\functionof{G}{r}}{r}$ is the gravitational force.  

In the hydrostatic limit, gravity dominates centrifugal terms; assuming $\pd{\rchi}{z} \lesssim \pd{\rchi}{\theta}$, the first term on the right side of Equation~\eqref{xDotGrad14} is \mbox{$\mathcal{O}\!\!\left(\!\tfrac{\lambda \Omega^2}{g}\!\right) \ll 1$} compared to the second.  The TWB equation is obtained by substituting Equation~\eqref{xDotGrad14} into $\vect{X}\leedot\grad$ of Equation~\eqref{reducedpde}:
	\begin{align}
	\vect{X} \leedot \grad \frac{{L^2}'}{2 \lambda^2} &= \vect{X} \leedot \grad {H}' - \frac{1}{\gamma-1} \vect{X} \leedot \grad \frac{p \sigma'}{\rho}  \nonumber\\
	\pd{\Omega^2}{z} &= \frac{g}{\gamma \lambda r} \pd{\sigma}{\theta} \left( 1 + \mathcal{O}\!\!\left(\!\tfrac{\lambda \Omega^2}{g}\!\right) \right). \label{eq:twb}
	\end{align}
Hence, we have derived the TWB equation by taking the appropriate limit of the G--S equation.

Note that there is no reference to stream surfaces in the final expression in Equation~\eqref{eq:twb}.  Indeed, the case of zero poloidal flow is a singular limit, as the equations that constrained $L$ and $\sigma$ to be constant along stream surfaces (e.g.,~$\vect{v}\leedot\grad\sigma = 0$) are trivially satisfied.  Thus, the two scalar variables, $\Omega$ and $\sigma$, are only required to satisfy the single Equation~\eqref{eq:twb}, and the problem is underdetermined.  

\citet{balbus2009simple} considered closing this TWB model by assuming the alignment of rotation and entropy contours (i.e.,~$\sigma = \functionof{\sigma}{\Omega}$), thus obtaining a system where a choice of the rotation and entropy on the surface can be uniquely extrapolated into the interior.  This assumption is strikingly similar to the appearance of stream surfaces in the G--S formalism (when $|\vect{v}_\textrm{p}| > 0$); since $L$ and $\sigma$ are both functions of $\rchi$, this implies that $\sigma = \functionof{\sigma}{L}$.  It should be noted that \citet{balbus2009simple} considered this \mbox{$\sigma$--$L$} alignment as well, but subsequently focused only on the \mbox{$\sigma$--$\Omega$} assumption due to a perceived stronger qualitative agreement with CZ observations.  However, as shown in Figure~\ref{fig:balbusrotationcomparison}, this apparent advantage arises from the difference in the choice of the entropy profile at the surface, not from the effect of the \mbox{$\sigma$--$\Omega$} alignment.  When the two choices of entropy closure are put on equal footing (i.e.,~with the same rotation and entropy at the surface), we find that both do equally well in capturing the characteristic rotation in the CZ.

Thus, we consider this relatively unexplored \mbox{$\sigma$--$L$} alignment as the natural limit of axisymmetric fluid equilibrium with vanishing poloidal flows, and using this framework, we evaluate the effects of including a nonzero poloidal flow on the rotation profile.

\subsection{Comparison with observations}
We have demonstrated that we can recover the qualitative features of the CZ rotation profile with our G--S model, but what about quantitatively?  \citet{balbus2009differential} considered a more general form of rotation and entropy as a function of latitude, fitted the model to the value and tilt of the isorotation contours in the bulk of the CZ, and obtained a good agreement away from the surface and tachocline.  Likewise, by allowing the freedom for the stream functions, we obtain an equally striking fit to helioseismology data, as shown in Figure~\ref{fig:helioseismologyfit}.  

\subsection{Radial stratification and a modified entropy condition}
\citet{balbus2009differential} also introduced a slightly more general \mbox{$\sigma$--$\Omega$} connection by dividing the entropy into a spherically symmetric part and an aspherical part, positing that the latter ``residual entropy'' aligns with rotation, giving $\sigma = \functionof{\sigma_0}{r} + \functionof{\sigma_1}{\Omega}$.  Their motivations were twofold: first is the mathematically convenient fact that TWB (Equation~\eqref{eq:twb}) is insensitive to this radial entropy profile, so the form of their solution is unchanged; and second is that convection in the absence of rotation would be in dynamic equilibrium with a superadiabatic radial entropy gradient, so the aspherical effects of rotation should be considered with respect to this spherically symmetric state.  Indeed, simulations in \citet{miesch2006solar} found that isorotation contours are better aligned with this residual entropy than with the total entropy.  Now, we demonstrate how a similar entropy assumption can fit into the framework of the G--S equation.

\begin{figure*}
\begin{center}
\includegraphics*[trim={7.8cm 4.0cm 2cm 2.4cm},clip,width=8cm,angle=0]{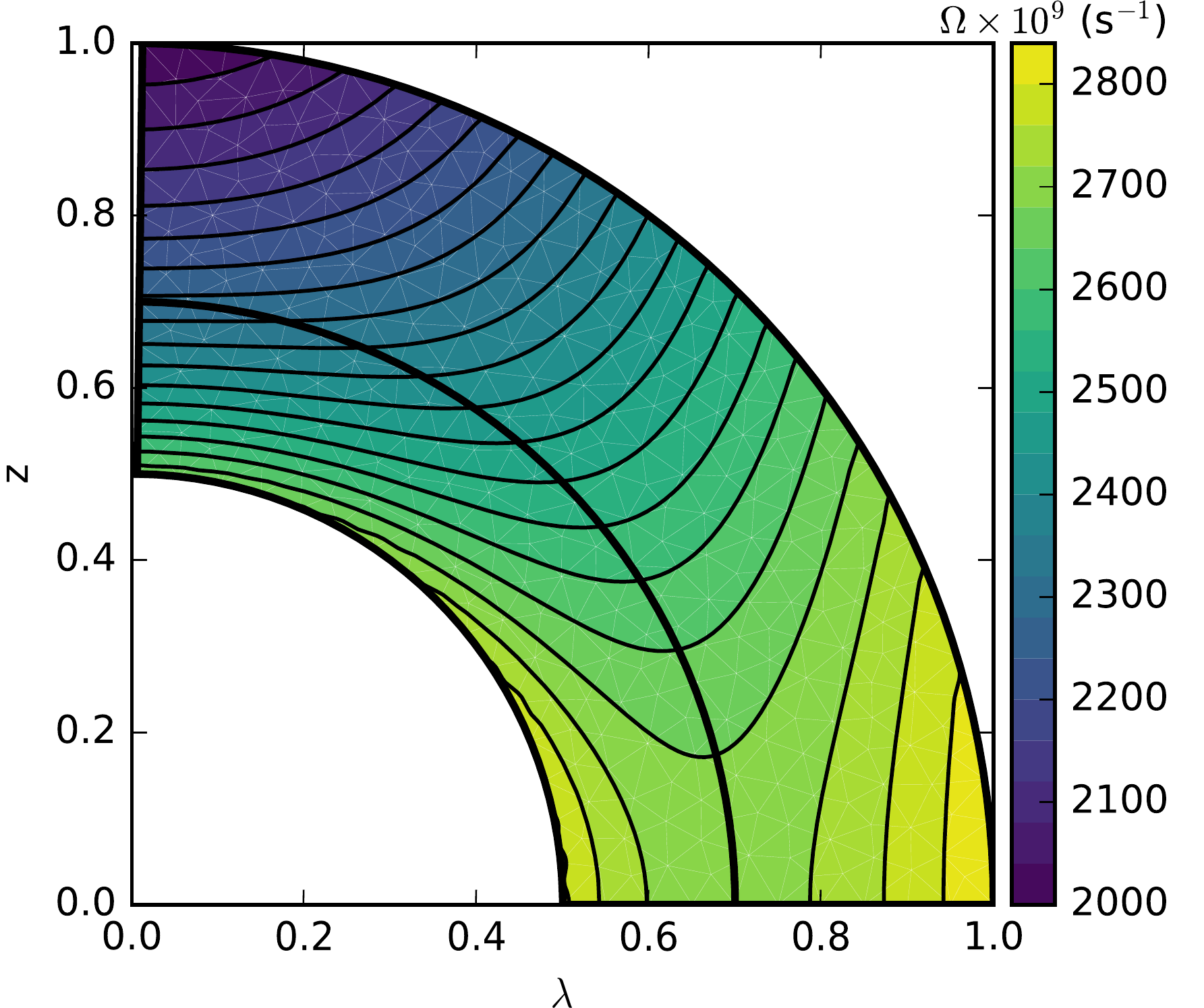}\includegraphics*[trim={7.8cm 4.0cm 2cm 2.4cm},clip,width=8cm,angle=0]{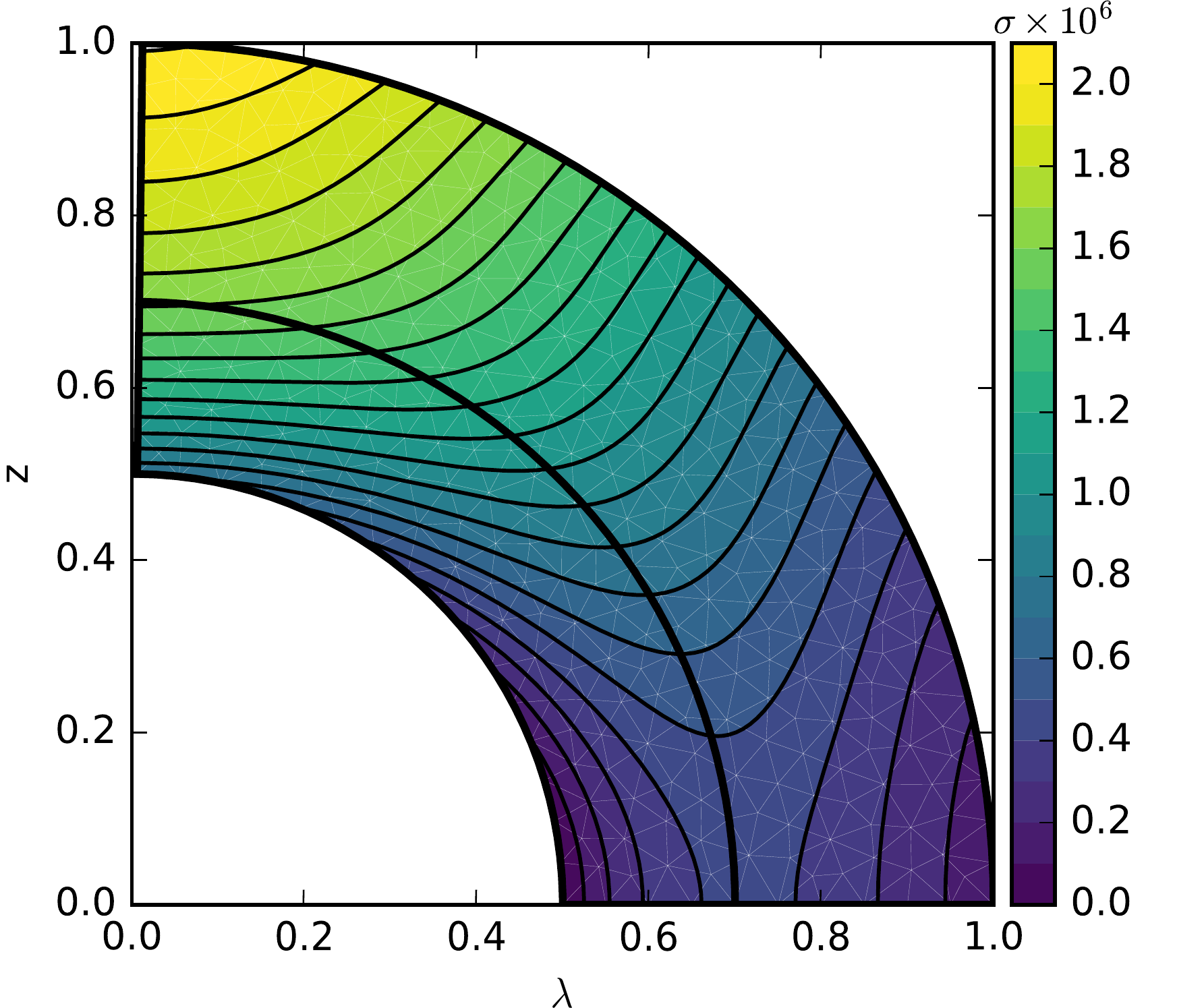}
\caption{Comparison of rotation and residual entropy in our G--S model.  \textbf{Left}: isorotation contours.  \textbf{Right}: total entropy contours \mbox{($\sigma = \functionof{\sigma_0}{r} + \functionof{\slashsigma}{\rchi}$).}  Note that the freedom afforded by the choice of radial entropy allows our model to emulate the $\sigma$--$\Omega$ alignment considered in \citet{balbus2009differential}.}
\end{center}
\label{fig:residualcomparison}
\end{figure*}

In the derivation of Equations~\eqref{goedbloedBernhydro} and~\eqref{goedbloedPDEhydro}, the assumption of stationary adiabatic flows implies that entropy is a stream function ($\vect{v}\cdot\grad \sigma = 0 \Rightarrow \sigma=\functionof{\sigma}{\rchi}$).  This assumption clearly does not hold in the Sun, where heat transport is dominated by radiation and turbulent convection (instead of mean-flow advection).  However, our motivations for using this adiabatic closure are likewise twofold: theoretical, because such a closure allows one to use the techniques of global MHD equilibrium and stability; and practical, because as we will see, the particular choice of closure matters much less than the overall latitudinal entropy profile (a free parameter in both models).  

Yet, as the radial entropy stratification is not small compared to the putative latitudinal gradient, it behooves us to separate out a spherically symmetric component.  Thus, in the spirit of \citet{balbus2009differential}, we assume the following form for the entropy:
	\begin{equation}
	\sigma = \functionof{\sigma_0}{r} + \slashsigma, \qquad \vect{v}\cdot\grad \slashsigma = 0,
	\end{equation}
so that the \textit{residual} entropy, $\slashsigma=\functionof{\slashsigma}{\rchi}$, is a stream function.  

We now trace the effects of this more general entropy assumption on the derivation of the Bernoulli equation~\eqref{goedbloedBernhydro} and G--S PDE~\eqref{goedbloedPDEhydro}.  For simplicity, we will outline the results for the HD limit, as the results are the same for the general MHD case.   The pressure term of Equation~\eqref{momentumeq} in the direction of $\vect{v}$ becomes
	\begin{align}
	\rho^{- 1} \vect{X} \leedot \grad p &= \rho^{- 1} \vect{X} \leedot \grad \! \left [ e^{\sigma_0 + \slashsigma } \rho^\gamma\right ] \nonumber\\
	&= -\frac{1}{\gamma - 1} \frac{p}{\rho} \vect{X} \leedot \grad \sigma_0 + \vect{X} \leedot \functionof{\grad}{\frac{\gamma}{\gamma - 1} \frac{p}{\rho}}.  \nonumber
	\end{align}
The second term yields the same entropy term as in Equation~\eqref{goedbloedBernhydro}, while the first term can be approximately incorporated by modifying the radial gravitational potential:
	\begin{align}
	\underbrace{\frac{1}{2}  \frac{\left | \grad\rchi \right |^2}{\rho^2\lambda^2} + \frac{1}{2} \frac{{L}^2}{\lambda^2} + \frac{\gamma \rho^{\gamma-1}}{\gamma-1} e^\sigma + G_\textrm{m} - H}_{\text{modified Bernoulli equation}} &= \underbrace{\int\frac{\delta T}{\gamma-1}  \vect{X} \leedot \grad \sigma_0 \frac{dl}{\left | \vect{X} \right |}}_{\text{$\mathcal{O} (\sigma_0 \delta T)$}}, \label{bernoulli2}
	\end{align}
where the integral is taken along the relevant stream surface, the modified gravitational potential is $\functionof{G_\textrm{m}}{r} = \functionof{G}{r} - \frac{1}{\gamma - 1} \int^r \! \frac{p_0}{\rho_0} \dd{\sigma_0}{r} dr$, and the temperature perturbation is $\delta T \equiv p/\rho - p_0/\rho_0$.  The quantities $p_0$ and $\rho_0$ are the background radial pressure and density profiles, which are given by hydrostatic balance:
	\begin{equation}
	\functionof{\rho_0}{r} = \smallparenth{\frac{\gamma-1}{\gamma} \frac{\left < H \right > - \functionof{G}{r}}{e^{\sigma_0(r)}}}^{\!\!\frac{1}{\gamma-1}}, \label{densitystrat}
	\end{equation}
where $\functionof{p_0}{r} = \rho_0^\gamma e^{\sigma_0}$ and angled brackets denote the mean value (note that $H-\left < H \right > \ll \left < H \right >$ in the solar case).

In obtaining the PDE~\eqref{goedbloedPDEhydro}, the pressure and gravity terms in the momentum equation~\eqref{momentumeq} are canceled by terms from $\rho \grad$ of Equation~\eqref{goedbloedBernhydro}.  Consider the corresponding terms in $\rho\grad$ of Equation~\eqref{bernoulli2}:
	\begin{align}
	\rho\grad H &= \rho \grad G_\textrm{m} + \frac{\gamma}{\gamma-1} \rho \functionof{\grad}{\frac{p}{\rho}} + \cdots \nonumber\\
	&= \!\!\!\!\!\!\underbrace{\vphantom{\frac{1}{\gamma-1}} \rho \grad G + \grad p}_{\text{cancel with Equation~\eqref{momentumeq}}} \!\!\!\!+\,\,\,\, \underbrace{\frac{1}{\gamma-1} p \grad\slashsigma}_{\text{term in PDE \eqref{goedbloedPDEhydro}}} \,\,\,\,+\,\,\,\, \underbrace{\frac{1}{\gamma-1} p \frac{\delta T}{T} \grad \sigma_0}_{\text{$\mathcal{O}\!\!\left(\sigma_0 \tfrac{\delta T}{T} p\right)$}} + \cdots. \nonumber
	\end{align}
Thus, both the PDE and the Bernoulli equation are changed by terms of size $\mathcal{O}\!\!\left(\sigma_0 \tfrac{\delta T}{T}\right)$, which is much smaller than the dominant balance of size $\mathcal{O}(\slashsigma)$ between differential rotation and residual entropy.  Indeed, the radial entropy profile is only weakly superadiabatic in the CZ; from mixing-length arguments, the change in entropy over a pressure scale height is ${\rm\sim~\!\!10^{-7}}$ at the base and ${\rm\sim~\!\!10^{-2}}$ at ${\rm\sim~\!\!0.5\%}$ below the surface \citep{lantz1999anelastic}, so $\slashsigma \lesssim \sigma_0 \ll 1$.  If differential rotation is driven primarily by baroclinic forcing, then $\tfrac{\delta T}{T} \sim \slashsigma \ll 1$, so this $\mathcal{O}\!\!\left(\sigma_0 \tfrac{\delta T}{T}\right)$ term can be dropped.  However, this term may not be negligible in the radiative interior (where $\sigma_0$ is strongly subadiabatic) or at the surface (where $\rho \rightarrow 0$).  Note that both our model and Balbus's break down in these regions.

What have we gained by demonstrating the insensitivity of our model to a radial entropy profile?  Using data from the 3D CZ simulations of \citet{miesch2006solar}, \citet{balbus2009differential} subtracted a particular choice of radial profile to show that the residual entropy can be made to align with rotation.  Here, we show that a different choice can be made to align it with angular momentum.  To demonstrate this, we add a radial profile that aligns our model's total entropy with rotation, as seen in Figure~\ref{fig:residualcomparison}.  Thus, by subtracting this radial profile, Balbus's relationship between rotation and residual entropy can be transformed into a relationship between angular momentum and a different choice of residual entropy.

\section{Effects of a Weak Poloidal Flow}
\label{sec:poloidaleffects}
\begin{figure*}
\begin{center}
\includegraphics*[trim={5.0cm 0 0 0},clip,width=5cm,angle=0]{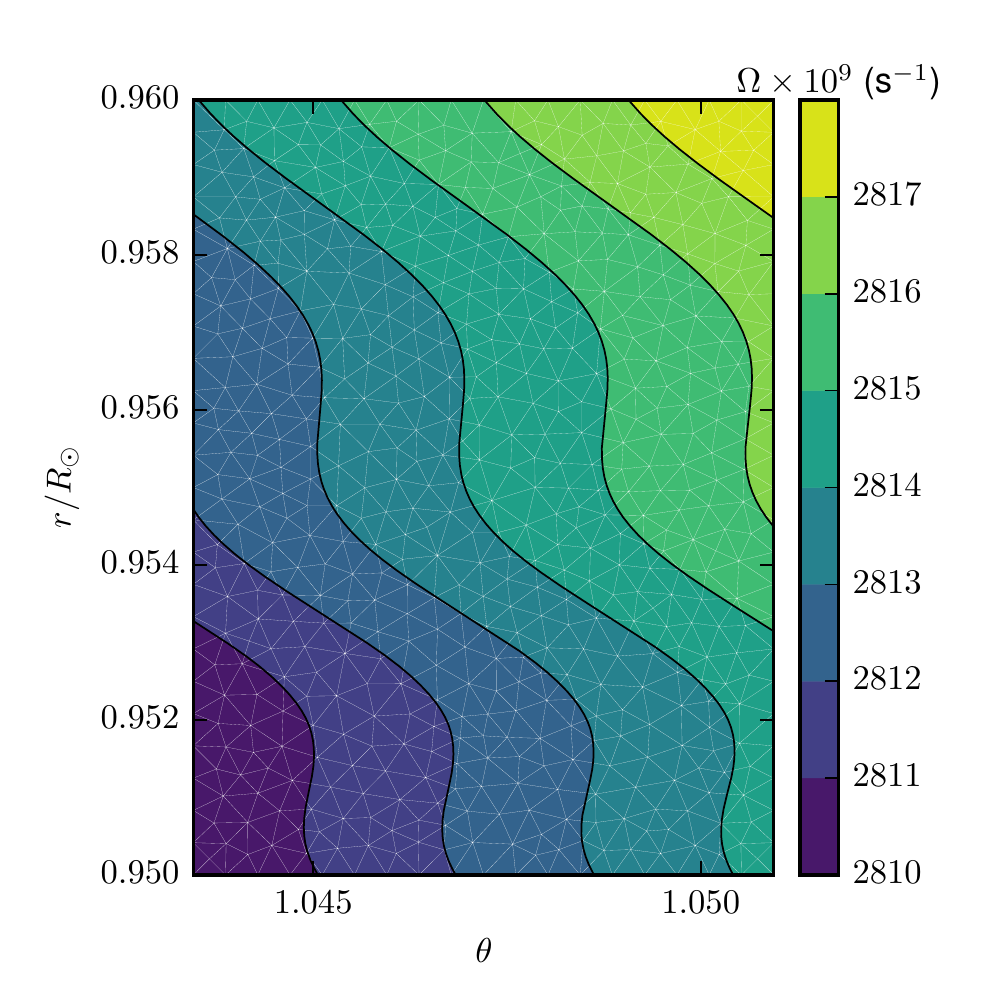}
\includegraphics*[trim={5.0cm 0 0 0},clip,width=5cm,angle=0]{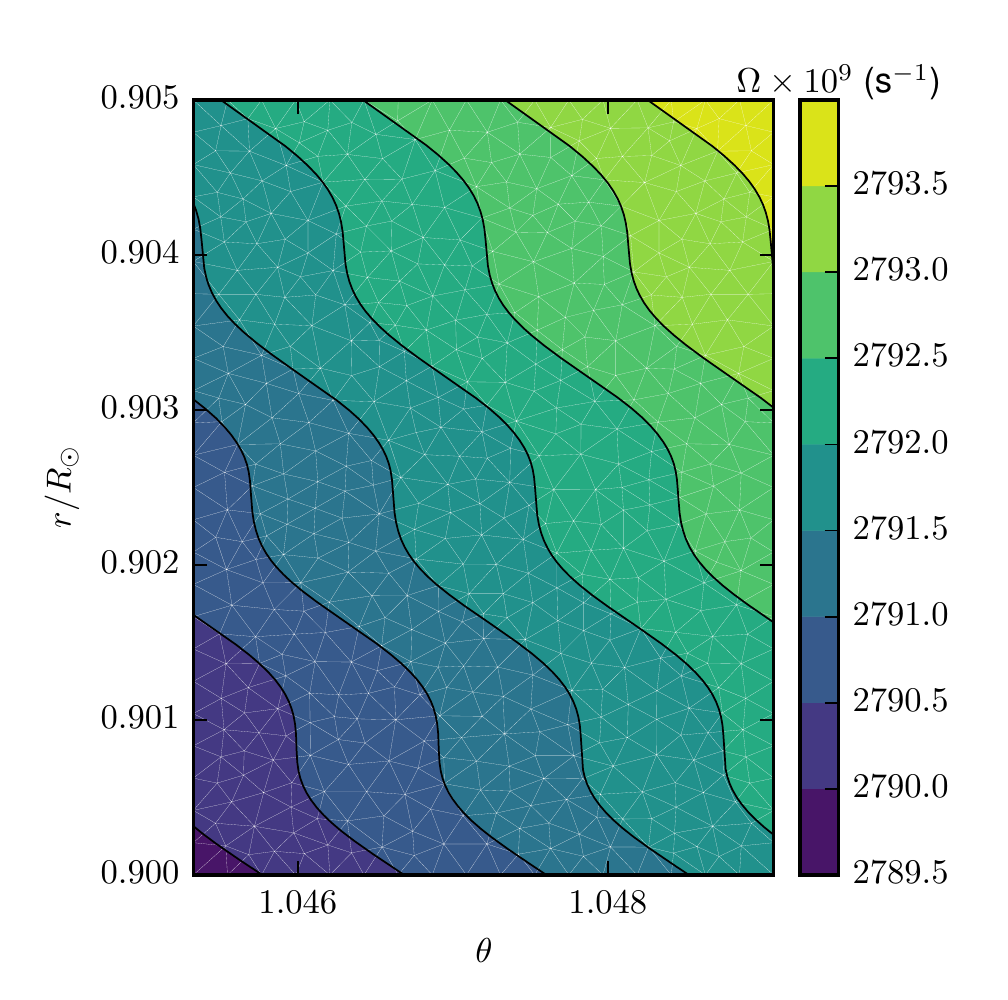}
\includegraphics*[trim={5.0cm 0 0 0},clip,width=5cm,angle=0]{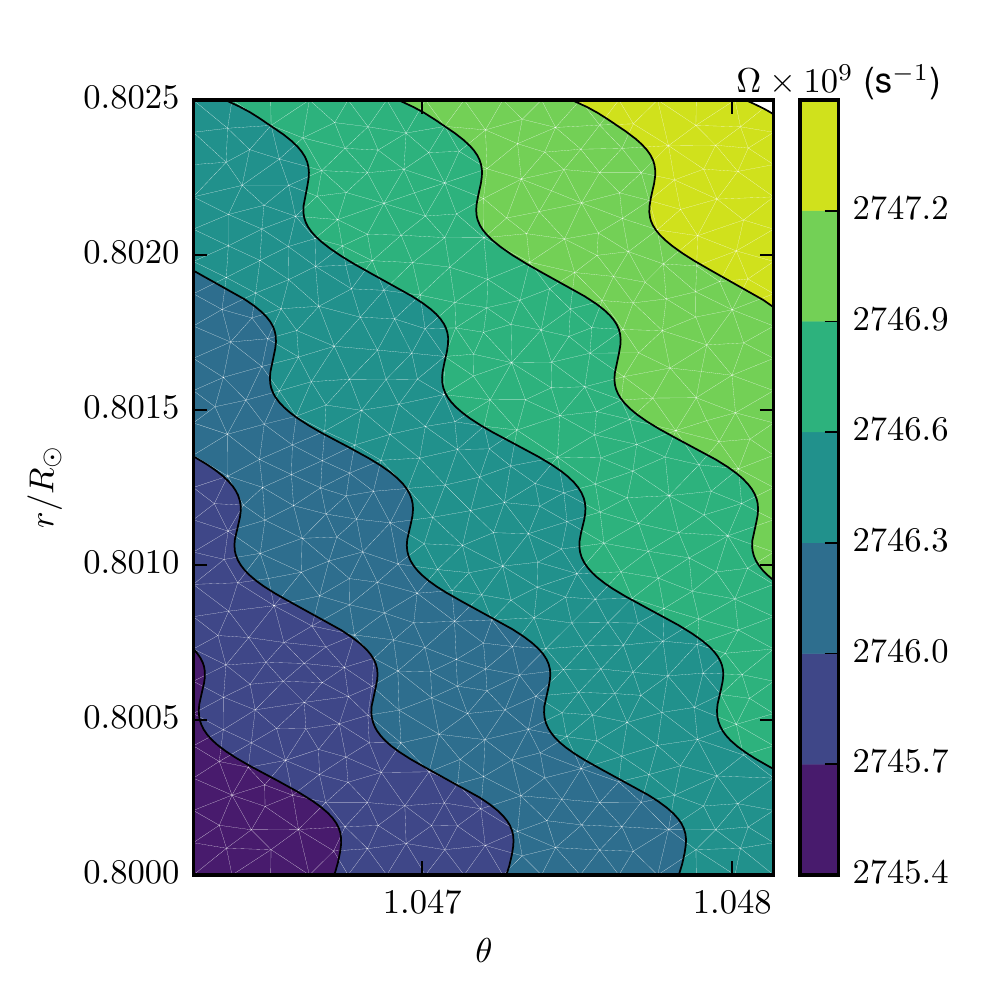}
\caption{Epicyclic oscillations around the smooth TWB solution in the CZ.  A perturbation to the gradient is applied to the upper boundary, and nonreflecting boundary conditions (see Appendix~\ref{sec:bc}) are applied to the sides and bottom.  The figures display isorotation contours in a small section at $30^\circ$ above the equator at decreasing radii from left to right: $r/R_\odot$~$=$~$0.95, 0.9, 0.8$.  Note the size of the domain; the perturbations have a short \change{wavelength, in} agreement with Equation~\eqref{oscillationLimit}.}
\end{center}
\label{fig:czoscillations}
\end{figure*}

As the G--S model requires a nonzero poloidal flow to motivate the alignment of entropy and angular momentum, we now investigate its effects on the TWB solutions considered in Section~\ref{sec:balbus}.  Since the poloidal flows are weak compared to the toroidal flows, we expect the poloidal contributions to the momentum equation (i.e.,~the differential part in Equation~\eqref{schematicequation}) to be small compared to the other terms (i.e.,~the algebraic part).  Indeed, the ratio of these terms is $ \vect{v}_\textrm{p}^2 / v_\phi^2 \lesssim10^{-4}$, so we expect the algebraic part to be the dominant term in the bulk of the CZ.  However, near the surface, the decreasing density along with conservation of mass flux implies an increase in the poloidal flow.  Thus, the differential part associated with it affects the equilibrium solution at leading order.  In this section, we show that deviations from the TWB solution due to the poloidal flow appear as short wavelength oscillations in the bulk of the CZ and as a slowing of rotation near the surface.

\subsection{Small oscillations in the convection zone}

In Section~\ref{sec:balbus}, we neglected the poloidal terms, leading to an algebraic equation that uniquely determined the solution, thus there was no need for boundary conditions.  However, when including poloidal terms, the resulting PDE specifies a continuum of possible solutions, corresponding to different choices of boundary conditions.  As the differential operator is subdominant, we linearize about the TWB solution and search for time-independent ($\omega = 0$) perturbations, giving a family of nearby equilibria.

First, we fix a background hydrostatic density profile, $\rho_0$.  As the TWB \change{solution, $\rchi_\textrm{TWB}$, satisfies} \mbox{$\functionof{F}{\rchi_\textrm{TWB}, \rho_0, \lambda, z} =0$,} we consider a nearby smooth solution, $\rchi_0$, that satisfies Equation~\eqref{schematicequation},
	\begin{equation}
	\deltastar \!\left [ \rchi_0 \right] + \functionof{F}{\rchi_0, \rho_0, \lambda, z} =0. \nonumber
	\end{equation}
As the scale length of the solution $\rchi_\textrm{TWB}$ is similar to the size of the CZ, the differential operator is indeed a small parameter, and $\rchi_0$ will not be significantly different from $\rchi_\textrm{TWB}$.  

However, if we consider a short wavelength perturbation to this solution, the differential operator can become comparable.  The linearized equation for a perturbation $\rchi_1 = \rchi - \rchi_0$ is
	\begin{equation}
	\delta\!\deltastar \!\left [ \rchi_0 ; \rchi_1 \right] + \rchi_1\left.\pd{F}{\rchi}\right|_{\rchi=\rchi_0}=0, \label{linearizedPDE}
	\end{equation}
where $\delta\deltastar\!\left [ \rchi_0 ; \rchi_1 \right]$ is the functional derivative of $\deltastar$ with respect to $\rchi_1$, evaluated at $\rchi_0$.  Let $\rchi_1$ be a local perturbation of the form $e^{i \vect{k}\cdot\vect{x}}$, so that $\grad \rchi_1 \sim i \vect{k} \rchi_1$ and $\nabla^2 \rchi_1 \sim - k^2 \rchi_1$.  By keeping only terms with at least a power of $\vect{k}$, the linearized differential operator becomes
\begin{align}
\delta\!\deltastar \!\left [ \rchi_0 ; \rchi_1 \right] &\approx \frac{k^2}{\rho\lambda^2} \rchi_1 - i \vect{k} \leedot \functionof{\grad}{\frac{1}{\rho\lambda^2}} \rchi_1. \label{linearizedDifferential}
\end{align}
Equations~\eqref{linearizedPDE} and~\eqref{linearizedDifferential} give a local dispersion relation,
\begin{align}
k^2 + i \vect{k} \leedot \grad \functionof{\ln}{\rho\lambda^2} \approx \frac{ |\grad\rchi |^2 F'}{\rho \vect{v}_\textrm{p}^2} \equiv K^2. \label{dispersionRelation}
\end{align}
We identify two limiting cases of this relation:
\begin{align}
k^2 &\approx K^2, \qquad \left| \grad \functionof{\ln}{\rho\lambda^2} \right| \ll K, \label{oscillationLimit}\\
i \vect{k} \leedot \grad \functionof{\ln}{\rho\lambda^2} &\approx K^2, \qquad \left| \grad \functionof{\ln}{\rho\lambda^2} \right| \gg K. \label{exponentialLimit}
\end{align}

Let us evaluate $K^2$ in the solar regime.   To obtain $F'$, we note that the stream surfaces are nearly cylindrical $\left(\rchi \approx \functionof{\rchi}{\lambda} \right)$, and differentiate $F = 0$ with respect to $\rchi$ at constant $\rho$:
\begin{align}
F' &\approx - \pd{F}{\lambda} \dd{\lambda}{\rchi} = \rho \dd{\lambda}{\rchi}\frac{{L^2}'}{\lambda^3} \nonumber\\
 &\approx \frac{\rho}{\lambda^3} {\smallparenth{\dd{\lambda}{\rchi}}\!}^2 \functionof{\dd{}{\lambda}}{\lambda^4 \Omega^2}. \nonumber
\end{align}
Then $K^2$ is given by
\begin{align}
K^2 \approx \frac{1}{\vect{v}_\textrm{p}^2}\frac{1}{\lambda^3}\functionof{\dd{}{\lambda}}{\lambda^4 \Omega^2}. \label{TPK2}
\end{align}
In the Sun, $K^2 > 0$, and (in the regions away from the surface and rotation axis) the first limit (Equation~\eqref{oscillationLimit}) is applicable.  Thus, in the bulk of the CZ, deviations from TWB take the form of sinusoidal perturbations around $\rchi_{0}$, with a wavelength that is inversely proportional to the density \mbox{($k^{- 1} \sim K^{- 1} \propto |\vect{v}_\textrm{p}| \propto \rho^{- 1}$).}  Equation~\eqref{TPK2} suggests an interpretation of these perturbations as the epicyclic motion of a fluid element, traced out in space via advection by the poloidal flow.

Numerically, we first solve for the smooth TWB solution in a small section of the CZ, then add a perturbation to the gradient on the upper boundary.  Figure~\ref{fig:czoscillations} shows the results for three different sections of the CZ, with increasing depths, demonstrating that the wavelength of the perturbations decreases according to Equation~\eqref{TPK2}.  As the wavelength of these perturbations is small (${\rm\sim~\!\!0.001R_\odot}$ at $r = 0.9R_\odot$), any significant deviation from the smooth solution in the CZ would result in large amplitude, oscillatory gradients in the rotation.  Even if these structures could be resolved by observations, they are likely to be suppressed (e.g.,~due to turbulent viscosity from convective motions).  Thus, in the bulk of the CZ, we expect to see the smooth TWB solution, and in Section~\ref{sec:EquilibriumNSSL}, we choose boundary conditions that minimize these oscillatory gradients.

\subsection{Equilibrium near surface shear layer}
\label{sec:EquilibriumNSSL}
\begin{figure*}
\begin{center}
\includegraphics*[trim={0 1.1cm 0 2.0cm},clip,width=7.25cm,angle=0]{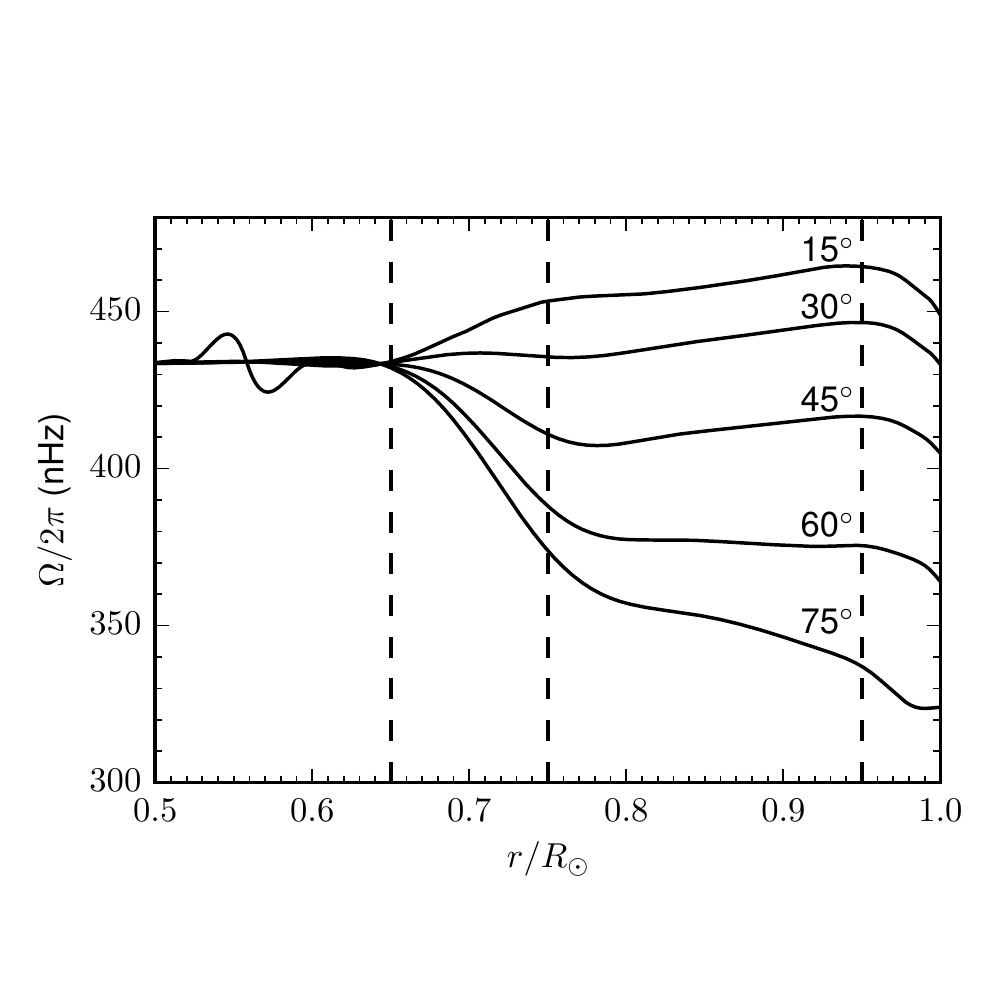}\includegraphics*[trim={0 1.1cm 0 2.0cm},clip,width=7.25cm,angle=0]{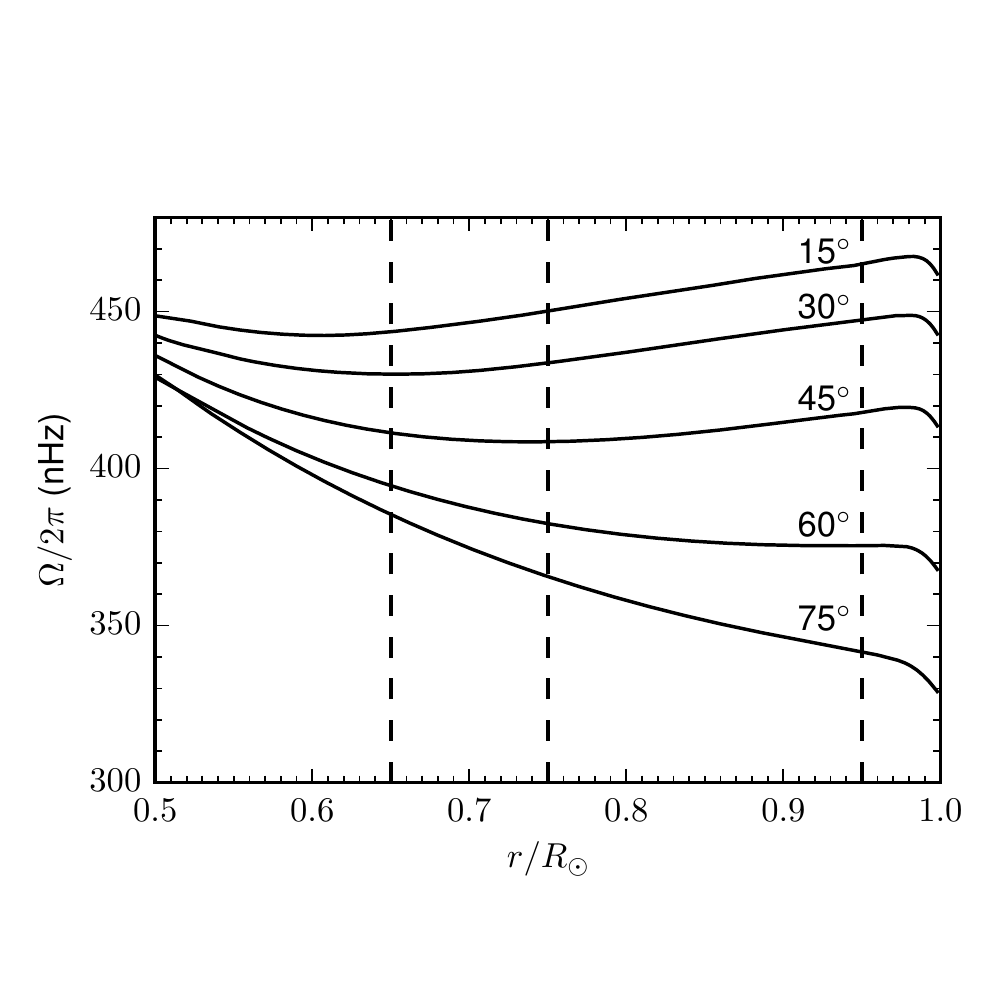}
\caption{\textbf{Left:} rotation as a function of radius for several latitudes, as obtained by helioseismology \citep{howe2011first}.  Dashed lines at $0.65R_\odot$, $0.75R_\odot$, and $0.95R_\odot$ provide approximate divisions between the radiative interior, tachocline, CZ, and NSSL.  \textbf{Right:} the same plot for our HD G--S model, with ${\rm\rho = 1\,~kg~m^{-3}}$, and ${\rm v_p = 20~m~s^{- 1}}$ at $r=0.99R_\odot$.  The profile shows a decrease in rotation near the surface that is qualitatively similar to the NSSL, but occurs in a region that is a factor of ${\rm\sim~\!\!3}$ smaller.  One notable feature missing from our profile is the transition to uniform rotation at the tachocline.  However, there are two reasons to believe that magnetic effects could become important at this layer and below: poloidal flows become weaker as the convective driving is suppressed in these stably stratified regions; and the effects of a putative interior fossil field \citep{gough1998inevitability} could enter the equilibrium at lowest order.}
\end{center}
\label{fig:nsslEquilibrium}
\end{figure*}

Near the surface, the differential part of the PDE~\eqref{goedbloedPDEhydro} becomes dominant, affecting the smooth solution at leading order.  Hydrostatic balance suggests the density decreases to zero as $\functionof{\rho}{r} \propto (R_\odot - r)^{\frac{1}{\gamma-1}}$.  Thus, $\grad \ln \rho$ diverges radially and $K^2 \rightarrow 0$, so we are in the second limit (Equation~\eqref{exponentialLimit}).  The dominant balance is now given by
	\begin{align}
	\pd{\rchi}{r} - \frac{K^2}{ \dd{}{r} \ln \rho} \rchi_1 \approx 0.
	\end{align}
Thus, $\pd{\rchi}{r} \rightarrow 0$, so the cylindrical stream surfaces in the CZ become oriented radially as they approach the surface.  As angular momentum is constant along stream surfaces, this corresponds to a slowing of the rotation at the surface (see Figure~\ref{fig:nsslEquilibrium}).  This transition occurs when the density scale height, $|\grad \ln \rho |^{- 1}$, becomes comparable to the wavelength given by Equation~\eqref{oscillationLimit}, which occurs in the outer ${\rm\sim~\!\!2\%}$ the solar surface.  While the effects of this transition are qualitatively similar to the NSSL, the location is confined much closer to the surface, so HD equilibrium is suggestive, but insufficient, as an explanation for the NSSL.

\section{Discussion}
\label{sec:conclusions}

In this paper, we employed the hydrodynamic limit of the Grad--Shafranov equation to obtain gravitationally dominated equilibrium states with a solar-like differential rotation.  In this limit, the surfaces of constant entropy align with those of angular momentum.  We compared results from our model to those with the entropy--rotation alignment in \citet{balbus2009simple} and found an equally good agreement with helioseismic observations.  Moreover, our model was able to reproduce qualitative features of the \change{near surface shear layer}.

Similar to \citet{balbus2009differential}, we demonstrated that it is possible to introduce a small radial entropy gradient while leaving the form of the equilibrium equations unchanged to leading order.  With the freedom afforded by this arbitrary radial profile, we showed that an aspherical entropy aligned with angular momentum can be changed into one that is nearly aligned with rotation.  Thus, the numerical simulations that demonstrate an alignment of aspherical entropy with rotation \citep{balbus2009differential} could just as well support an alignment with angular momentum by using a different choice of spherical entropy gradient. 

We then allowed for a weak poloidal flow and considered its effects on the equilibrium.  In the convection zone, perturbations around the smooth solution appear with a wavelength that is inversely proportional to the density.  These epicyclic oscillations are small compared to the scale length of the background solution (${\rm\sim~\!\!0.01R_\odot}$ at $r \sim 0.97R_\odot$, decreasing to ${\rm\sim~\!\!0.001R_\odot}$ at $r \sim 0.9R_\odot$), and boundary conditions were chosen to minimize them.  Thus, the addition of a solar-relevant poloidal flow does not significantly change the equilibrium profile in the bulk of the convection zone.  However, a decrease in the rotation appears within ${\rm\sim~\!\!2\%}$ of the solar surface, which is reminiscent of the near surface shear layer, albeit smaller.

One limitation of our model involves the topology of the meridional circulation.  Near the surface, its poloidal flow becomes radial, which is at odds with reasonable boundary conditions and perpendicular to the observed poleward flow.  Likewise, its axially aligned subsurface flows thread the entire interior and do not exhibit the expected recirculation at the tachocline or equator.  In fact, as angular momentum increases with cylindrical radius everywhere, recirculating flows are incommensurate with \mbox{solar-like} rotation within our model, as they are required to conserve angular momentum.  One possible resolution is that strong convection at the solar surface effectively ``rewrites'' the stream functions as the flow moves poleward.  Likewise, the strongly subadiabatic entropy gradient could serve a similar role at the base of the convection zone.

Future work will involve including a magnetic field.  In contrast to flow velocities, which decrease with depth, magnetic fields have an increasing effect on the equilibrium profile.  Indeed, magnetism has been suggested as essential to understanding the tachocline and radiative interior \citep{rudiger1997slender,rudiger2007structure,gough1998inevitability,wood2011sun}, and cyclic variations in the mean flows are likely intertwined with the solar magnetic cycle \citep{vorontsov2002helioseismic,rempel2007origin,beaudoin2013torsional}.

Furthermore, to assess the stability of these equilibria, they could be used as initial conditions for time-dependent simulations. Relevant features that remain intact could then be considered to have an equilibrium description.  Our framework could also be applied to other rotating magnetized objects, such as planetary cores, gas giants, and other stars.

\acknowledgments
This research is supported by the U.S. Department of Energy under contract No.~DE-AC02-09CH11466, and the National Science Foundation under grant Nos.~AGS- 1338944 and AGS-1460169.

\appendix

\section{Numerical Method}
\label{sec:method}
\subsection{Finite element formulation}
We employ a finite element method (FEM) numerical scheme to solve the weak form of the Bernoulli equation~\eqref{goedbloedBernhydro} and PDE~\eqref{goedbloedPDEhydro}.  To obtain the weak form of the PDE, we multiply it by a test function, $\functionof{\xi}{\lambda,z}$, and integrate the differential operator by parts:
\begin{align}
\int_V \xi \left( \grad\leedot \left ( \frac{\grad\rchi}{\rho\lambda^2} \right ) + F(\rchi, \rho, \lambda, z)\right) \,dV  = 0\nonumber\\
\int_S \frac{\xi}{\rho\lambda^2} \left(\hat{\bf{n}}\leedot\grad\rchi\right) dS - \int_V \frac{\grad \xi \leedot \grad\rchi}{\rho\lambda^2} dV + \int_V \xi F\,dV  = 0, \label{weakPDE} 
\end{align}
where $\hat{\bf{n}}$ is the outward unit vector normal to the surface.  If a function, $\rchi$, satisfies this relation for all test functions, $\xi$, then it is said to be a weak solution of the PDE.  Notice that this formulation allows for a discontinuous $\grad\rchi$, as compared to the original PDE, which might not be well defined in this case.  The weak form of the Bernoulli equation is \change{simply}
	\begin{align}
	\int_V \xi \left( \frac{1}{2} \frac{\left | \grad\rchi \right |^2}{\rho^2\lambda^2} + \frac{1}{2} \frac{{L}^2}{\lambda^2} + \frac{\gamma\rho^{\gamma-1} }{\gamma-1} e^\sigma + G - H \right) dV &= 0\nonumber\\
	\int_V \xi B(\grad \rchi, \rchi, \rho, \lambda, z) \, dV &= 0.
	\label{weakBern}
	\end{align}

We discretize the domain into an unstructured triangular mesh using the code GMSH \citep{geuzaine2009gmsh}.  The discretized function space for $\xi$, $\rho$, and $\rchi$ is the set of piecewise polynomials: analytic within any triangle and continuous across the boundaries.  An order $p$ polynomial over two variables has $(p+1)(p+2)/2$ degrees of freedom, hence this many nodes are assigned to each triangle (with nodes on the boundaries associated with multiple triangles).
The basis function, $\xi_i$, associated to node $i$ is the continuous, piecewise polynomial with the value $1$ at node $i$ and $0$ at all others \change{(in this paper, we use cubic basis functions).}

Discretizing $\functionof{\rchi}{\lambda, z}$ as $\sum_i x_i \xi_i$ (summation of repeated indices will be henceforth implicit), Equation~\eqref{weakPDE} becomes a matrix equation, $A_{i\!j} x_{\!j} + b_i = 0$, for the coefficients, $x_{\!j}$:
	\begin{align}
	\underbrace{\int_S \frac{\xi_i}{\rho\lambda^2} \left(\hat{\bf{n}}\leedot\grad \xi_{\!j} \right) x_{\!j} dS}_{\textrm{see Appendix}~\ref{sec:bc}} \;\underbrace{- \int_V \frac{\grad \xi_i\leedot \grad \xi_{\!j}}{\rho\lambda^2} x_{\!j} dV}_{A_{i\!j}x_{\!j}} \; \underbrace{+ \int_V \xi_i F\,dV}_{b_{i}}  &= 0.\label{discretizedPDE}
	\end{align}
Likewise, we discretize $\rho(\lambda,z)$ as $\sum_i y_i \xi_i$, and the Bernoulli equation becomes
	\begin{align}
	\underbrace{\int_V \xi_i B\,dV}_{c_{i}} = 0. \label{gaussBern}
	\end{align}
To approximate the integrals, we use Gaussian quadrature, e.g.,
	\begin{align}
	\sum_{n=1}^{N_\textrm{line}} \left[ \sum_{q} w_{qn}  \frac{\xi_i}{\rho\lambda^2} \left(\hat{\bf{n}}\leedot\grad \xi_{\!j} \right) x_{\!j} \right] + \sum_{n=1}^{N_\textrm{tri}} \left[ -\sum_{q} w_{qn} \frac{\grad \xi_i \leedot \grad \xi_{\!j}}{\rho\lambda^2} x_{\!j} + \sum_{q} w_{qn} \xi_i F \right] &= 0, \label{gaussPDE}
	\end{align}
where $N_\textrm{line}$ and $N_\textrm{tri}$ are the numbers of lines and triangles, respectively, and the parts inside the sums are evaluated at the $q$th quadrature point of element $n$ with the associated weight, $w_{qn}$.

\subsection{Boundary conditions}
\label{sec:bc}
In Equation~\eqref{gaussPDE}, the sum over the triangles contains the terms $A_{i\!j}x_{\!j}$ and $b_i$.  The treatment of the sum over the lines depends on the boundary conditions.  For Dirichlet boundary conditions, we simply fix the value of the nodes on the boundary by substituting the corresponding row of $A$ with that of the identity and changing the entry of $b$ to the desired boundary value.  For Neumann boundary conditions, $\hat{\bf{n}}\leedot\grad\rchi=\rchi'_\textrm{bc}$, we make precisely this substitution and include it in $b_i$:
	\begin{align}
	b_i \rightarrow b_i + \sum_{n=1}^{N_\textrm{line}} \left[ \sum_{q} w_{qn} \frac{\xi_i}{\rho\lambda^2} \rchi'_\textrm{bc} \right].
	\end{align}
These boundary conditions are appropriate when a physical constraint must be satisfied on the boundary (e.g.,~symmetry constraints at the poles and equator).  However, in cases where the boundaries of the computational domain are not physical (i.e.~when considering a subregion of the CZ, see Figure~\ref{fig:czoscillations}), we need an analog of nonreflecting boundary conditions.  In this case, we incorporate the surface term in Equation~\eqref{gaussPDE} into the matrix $A$:
	\begin{align}
	A_{i\!j} \rightarrow A_{i\!j} + \sum_{n=1}^{N_\textrm{line}} \left[ \sum_{q} w_{qn} \frac{\xi_i}{\rho\lambda^2} \hat{\bf{n}}\leedot\grad \xi_{\!j} \right].
	\end{align}
This amounts to ensuring that the discretized weak form of the PDE is satisfied everywhere within the domain, but may lead to a formally underdetermined system, as no explicit boundary conditions are imposed.  Hence, we introduce a regularization term to determine the system (see Appendix~\ref{sec:reg}).

\subsection{Iteration scheme}
\label{sec:iteration}
Once the weak form is discretized, Equation~\eqref{gaussPDE} essentially becomes
	\begin{align}
	\functionof{A_{i\!j}}{\rchi,\rho} x_{\!j} + \functionof{b_i}{\rchi,\rho} = 0. \nonumber
	\end{align}
We converge on a solution for $\rchi$ by iteratively solving the linearized problem, i.e.,~at the $n$th stage,
\begin{align}
\varepsilon_i^n &= A_{i\!j} x_{\!j}^n + b_i \nonumber\\
x_{\!j}^{n+1} &= x_{\!j}^{n} - \left(A_{i\!j}\right)^{\!-1} \varepsilon_i^n. \label{iterationScheme}
\end{align}
This scheme works well, provided that $A$ and $b$ only weakly depend on $\rchi$.  However, this may not be the case when the algebraic part of Equation~\eqref{gaussPDE} (i.e.,~the sum involving $F$) is the dominant term.  Thus, we incorporate its derivative as follows:
	\begin{align}
	x_{\!j}^{n+1} &= x_{\!j}^{n} - \left(A_{i\!j} + \pd{b_i}{x_{\!j}} \right)^{\!\!-1} \! \varepsilon_i^n. \label{iterationScheme2}
	\end{align}
Likewise, we update the density as follows:
	\begin{align}
	y_{\!j}^{n+1} = - \left(\pd{c_{i}}{y_{\!j}}\right)^{\!\!-1} \! c_i^{n}. \label{iterationSchemeBern}
	\end{align}

\subsubsection{Regularization}
\label{sec:reg}
This iteration scheme may lead to attempted inversion of a nearly singular matrix (e.g.,~the case of an underdetermined system due to nonreflecting boundary conditions), resulting in a large change in $x_{\!j}$, thus inhibiting convergence.  Hence, we introduce a regularization term.  In particular, at each step $n$, we minimize the following quantity:\change{
\begin{align}
\left| A_{i\!j}^n x_{\!j}^{n+1} - b_i^n \right|^2 + \alpha \left|x_{\!j}^{n+1} - x_{\!j}^n \right|^2,
\end{align}}
where $\alpha$ is a regularization parameter, which we choose as small as possible such that the change in $x_{\!j}$ is within some bound.  

In order for this regularization parameter to have a similar effect on \change{all the} nodes, we precondition the matrix as follows: 
	\begin{align}
	A_{i\!j} &\rightarrow \left(D_{ik}\right)^{-1} A_{k\!j}\nonumber\\
	b_i &\rightarrow \left(D_{ik}\right)^{-1} b_k\nonumber\\
	D_{ik} &= \left\{\begin{array}{ll}
		\sum_{\!j} |A_{i\!j}|, & \quad i=k\\
		0, & \quad \textrm{otherwise}
		\end{array}\right.
	\end{align}
Using this iteration scheme, we alternate solving Equations~\eqref{iterationScheme2} and~\eqref{iterationSchemeBern}, using the most recent values for $\rchi$ and $\rho$.

\subsection{Initial conditions}
\label{sec:ic}
It is beneficial to start with initial conditions that are close to the solution.  In our problem, the dominant terms in Equations~\eqref{goedbloedBernhydro} and~\eqref{goedbloedPDEhydro} are the gravitational part and algebraic part, respectively.  Thus, we first solve for $\rho^0$ and $\rchi^0$ at each node as follows:
	\begin{align}
	\rho^0:& \quad \text{hydrostatic} \nonumber\\
	\rchi^0:& \quad \functionof{F}{\rchi^0,\rho^0} = 0. \nonumber
	\end{align}
Then, we evaluate the differential parts (at the nodes) using a 2D spline interpolation, and iterate the following several times to obtain the initial conditions:
	\begin{align}
	\rho^n:& \quad \functionof{B}{\grad\rchi^{n-1},\rchi^{n-1},\rho^n,\lambda,z} = 0 \nonumber\\
	\rchi^n:& \quad \functionof{F}{\rchi^n,\rho^n,\lambda,z} = -\deltastar \!\left [ \rchi^{n-1} \right ]. \nonumber
	\end{align}

\end{document}